\documentclass[apjl]{emulateapj}
\usepackage{graphicx}
\usepackage{amsmath}  
\usepackage{amssymb} 
\usepackage[lofdepth,lotdepth,caption=false]{subfig}
\usepackage{natbib}
\usepackage{epsf}
\usepackage{enumitem}

\newcommand{\nstars}{four}

\newcommand{\andromeda}{$\text{M31}$}
\newcommand{\feh}{\ensuremath{\text{[Fe/H]}}}
\newcommand{\fehavg}{\ensuremath{\langle\text{[Fe/H]}\rangle}}
\newcommand{\afe}{\ensuremath{\text{[}\alpha/\text{Fe]}}}

\newcommand{\afeavg}{\ensuremath{\langle[\alpha/\text{Fe}]\rangle}}
\newcommand{\teff}{\ensuremath{T_{\text{eff}}}}
\newcommand{\logg}{\ensuremath{\text{log}\,g}}

\shorttitle{Alpha Abundances in \andromeda\ Halo}
\shortauthors{Vargas~et~al.}

\begin{document}

\title{\afe\, Abundances of Four Outer M\,31 Halo Stars}

\author{Luis C. Vargas\altaffilmark{1}}
\author{Karoline M. Gilbert\altaffilmark{2}}
\author{Marla Geha\altaffilmark{1}}
\author{Erik J. Tollerud\altaffilmark{1\,\dag}}
\author{Evan N. Kirby\altaffilmark{3}}
\author{Puragra Guhathakurta\altaffilmark{4}}

\affil{\altaffilmark{1}Department of Astronomy, Yale University, 260 Whitney Ave., New Haven, CT~06511, USA; luis.vargas@yale.edu\\
\altaffilmark{2}Space Telescope Science Institute, 3700 San Martin Dr., Baltimore, MD~21218, USA\\
\altaffilmark{3}California Institute of Technology, 1200 E.\ California Blvd., MC 249-17, Pasadena, CA 91125, USA\\
\altaffilmark{4}UCO/Lick Observatory and Department of Astronomy and Astrophysics, University of California, 1156 High St., Santa Cruz, CA 95064, USA}
\altaffiltext{\dag}{Hubble Fellow}

\begin{abstract}

We present alpha element to iron abundance ratios, \afe, for four stars in the outer stellar halo of 
the Andromeda Galaxy (M\,31). The stars were identified as high-likelihood field halo stars by 
\citet{Gilbert2012a} and lie at projected distances between 70 and 140 kpc from \andromeda's 
center. These are the first alpha abundances measured for a halo star in a galaxy beyond the 
Milky Way. The stars range in metallicity between \feh$ = -2.2$ and \feh$ = -1.4$. The sample's 
average \afe\ ratio is $+0.20\pm0.20$. The best-fit average value is elevated above solar which is 
consistent with rapid chemical enrichment from Type~II supernovae. The mean \afe\ ratio of our \andromeda\ outer 
halo sample agrees (within the uncertainties) with that of Milky Way inner/outer halo stars that have 
a comparable range of \feh.
\end{abstract}

\keywords{galaxies: abundances ---
          galaxies: individual (M31) ---
          galaxies: evolution ---
          Local Group}

\section{Introduction}\label{intro_sec}

\setcounter{footnote}{0}

The assembly of galactic stellar halos via accretion of substructure
is the defining feature of hierarchical galaxy formation \citep[e.g.,
][]{Searle1978a,White1978a}.  Stellar halos provide a direct link
between the present-day properties of a galaxy and the properties of
its cosmological progenitors.  In this \textit{Letter}, we focus on
chemical abundances of stellar halos to investigate this connection
for \andromeda.

Galaxy-scale $\Lambda$CDM simulations suggest 
that the accretion process deposits halo stars at all 
galactocentric distances \citep{Johnston2008a,Cooper2010a}.
In the inner halo, an additional component 
formed in-situ adds to the complexity of stellar halos 
\citep{Tissera2012a,Zolotov2012a}. In contrast, the 
outer regions of halos ($R\,\gtrsim{20}$ kpc) 
are exclusively formed by accretion; stars in 
the outer halo encode both properties of the 
halo's progenitors and when they were accreted 
\citep{Johnston2008a,Cooper2010a}.

We focus here on the alpha to iron abundance ratio, \afe\footnote{We
adopt the definition of \afe\ as the unweighted average of [Mg/Fe],
[Si/Fe], [Ca/Fe], and [Ti/Fe].}, of halo stars. Higher \afe\ values
are positively correlated with high star-formation rates and/or short
star formation histories \citep{Tinsley1979a}.  Higher \afe\ indicates
a galaxy's metal enrichment is dominated by alpha element-rich Type~II
supernovae (SNe), which have relatively short-lived progenitors, while
later Type~I SNe gradually lower \afe\ by depositing iron into the ISM
\citep[e.g.,][]{Woosley1995a,Nomoto2006a}.  Hence, determining \afe\
for outer halo stars constrains the star formation histories of the
halo's progenitors.

The Milky Way (MW) halo has been investigated with these 
ideas in mind.  It is composed of mostly metal-poor (i.e., 
low \feh) stars \citep{Ivezic2008a,Carollo2010a} 
with elevated \afe\,$\sim{+0.3}$ abundance ratios 
\citep[e.g.,][]{Fulbright2000a,Cayrel2004a,Cohen2004a}.
While most abundance measurements target nearby and 
thus inner halo stars, kinematical criteria have been used 
to identify outer halo stars passing through the inner 
halo for chemical abundances analysis. One exception is 
the study by \citet{Lai2009a}, who measured alpha abundances 
for a single star in the outer halo with a Galactocentric distance 
of $\sim{40}$ kpc. At the low metallicities characteristic of the 
outer halo \citep[\feh\,$\lesssim{-2}$;][]{Carollo2010a}, the stellar 
population also appears to be enhanced in \afe\ 
\citep[e.g.,][]{Roederer2009a,Ishigaki2012a}. This pattern is different 
from that of present day MW satellites, which 
have a significant fraction of low alpha abundance ratios (\afe\,$\lesssim{0}$) 
at \feh\,$\gtrsim{-2}$. Thus, \afe\ measurements suggest that stellar halo 
build-up was dominated by substructure with a chemical enrichment 
history different from the present day MW dwarf satellites. 

\begin{figure*}[tpb!]
\centering
\includegraphics[width=.99\textwidth]{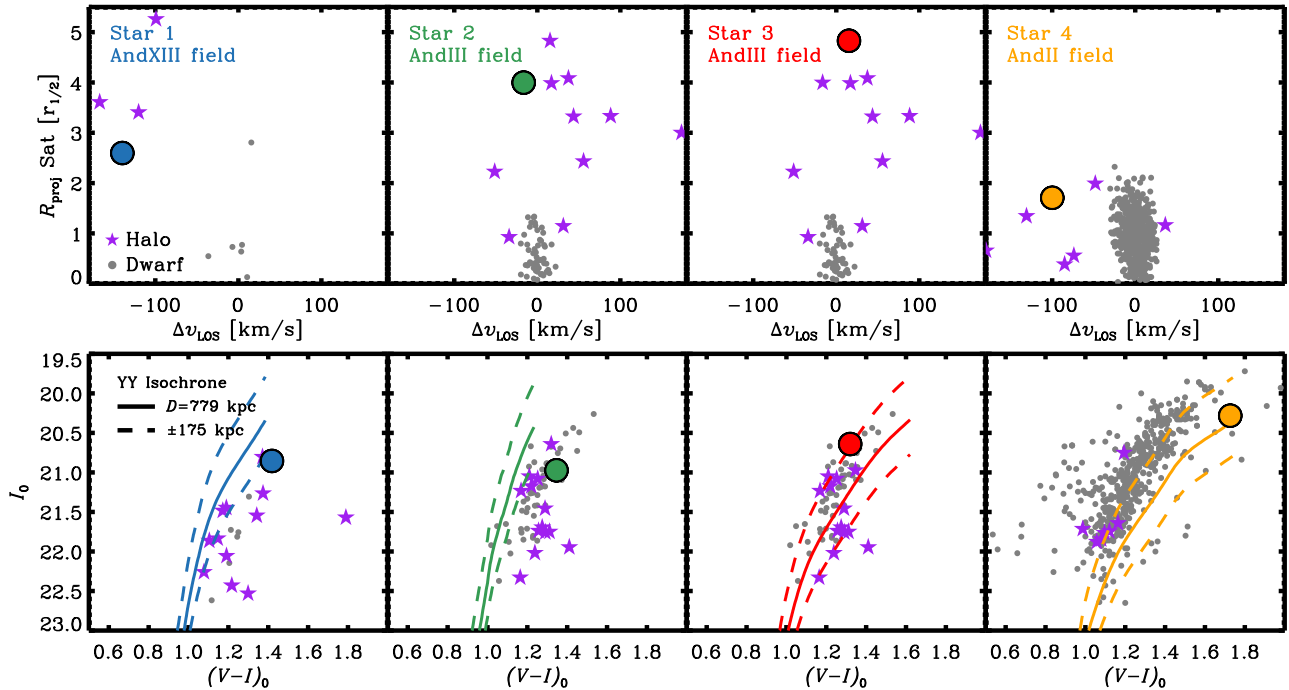}	
\caption{\textit{Top:} Velocity separation, $\Delta$$v_{\text{LOS}}$, versus
projected radius, $R_{\text{proj}}$, for each of the four field halo 
stars analyzed in this paper compared to the satellite galaxy located 
in its vicinity. Program stars are shown as large circles, and the same color 
coding is used throughout the paper. Other high-likelihood halo 
members from G12  are shown as purple 
stars. Satellite member stars (small grey circles) 
cluster at ($\Delta$$v$, $R_{\text{proj}}$)~$\sim(0,0)$. \textit{Bottom:} ($V-I_{0}$, $I_{0}$) 
CMD showing the same stars as in the 
top panels. We overlay 12\,Gyr Yale-Yonsei isochrones
\citep{Kim2002a} with \feh\ and \afe\ nearest to that measured 
spectroscopically for each of the program stars. We place the 
isochrones at the line-of-sight distance to \andromeda\ (solid lines) 
and at distances $\pm{175}$ kpc smaller/larger (dashed lines).}
\label{fig: membership}
\vskip .5pt
\end{figure*}

Simulations show a wide range in the accretion characteristics of
different halos \citep{Johnston2008a,Cooper2010a}, motivating
observational studies beyond the Galaxy. The \andromeda\ system is the
nearest massive galaxy and provides hints of a quite
different formation history relative to the MW.  It hosts more
satellite galaxies than the MW \citep{Martin2013b}, with a wider range
of chemical abundances \citep[][hereafter V14]{Ho2014a,Vargas2014a}.
\andromeda's halo lacks a characteristic density break
\citep[][hereafter G12]{Ibata2014a,Gilbert2012a}, which suggests an
extended accretion history \citep{Deason2013a}.  It also peaks at a
\feh\ value $> 1$ dex higher than the MW, qualitatively consistent
with a larger fraction of halo stars coming from more luminous (and
hence more metal-rich) satellites.

The metallicity-driven differences between the MW and \andromeda\ are
less clear in their outer stellar halos, however, because the average
stellar metallicity of \andromeda\ decreases with projected radius out
to more than 100 kpc
\citep{Kalirai2006a,Koch2008b,Ibata2014a,Gilbert2014a}.  This suggests
an underlying variation in metallicity of the progenitors of the
\andromeda\ stellar halo and calls for additional chemical probes
beyond $\feh$. However, stellar abundances other than \feh\ in the
\andromeda\ system have so far been limited to young supergiants in
the disk \citep{Venn2000a, Trundle2002a}. This \textit{Letter}
presents the first \afe\ ratios measured in \andromeda's stellar halo.

\section{Assembling The Sample}\label{sec_data}

\subsection{\citet{Gilbert2012a}'s Halo Membership}\label{ssec_membership}

G12 identified over 1600 \andromeda\ halo stars as part of the 
Spectroscopic and Photometric Landscape of Andromeda's 
Stellar Halo (SPLASH) survey. They used a combination of five 
diagnostics to calculate the relative likelihood of membership in 
the \andromeda\ halo versus a foreground MW population \citep{Gilbert2006a}. 
These diagnostics utilize photometric and spectroscopic measurements: 
the star's radial velocity, its position in a $(V-I, I)$ color-magnitude diagram
(CMD), its magnitude in the narrow-band DDO51 filter, the strength of the \ion{Na}{1} 
absorption line at 8190 \AA\, and the difference between photometric 
and calcium triplet (CaT)-based metallicities.  Stars are designated as 
\andromeda\ stars if it is more probable they are red giant branch (RGB)
stars at the distance of \andromeda\ than MW foreground stars.  Stars 
are designated to have a high likelihood of \andromeda\ membership if 
it is 3 times more likely that they are \andromeda\ RGB stars rather than 
MW stars. In Section~\ref{ssec_sample}, we additionally assess whether
the stars are \andromeda\ halo or \andromeda\ dwarf satellite members.

\subsection{Higher S/N Observations}\label{ssec_data}

To measure metallicities and alpha abundances, we obtained 
higher S/N spectra of halo stars from the sample described in 
Section~\ref{ssec_membership}. The targets were selected 
due to their relative proximity to various \andromeda\ satellites 
studied by V14, enabling us to observe them using the multi-object 
masks designed for the satellite targets.

The Keck/DEIMOS spectra span the wavelength range 
$6300 < \lambda < 9100$ \AA\ and have a resolving power of 
$\lambda/\Delta\lambda\sim{6000}$. The data reduction for the additional 
spectra follows that of the G12 sample, and was performed 
with a modified version of the spec2d/DEEP pipeline \citep{Cooper2012a,
Newman2013a} adapted to stellar spectra \citep{Simon2007a,Kalirai2010a}.  

Including the new observations, we identified ten likely \andromeda\ 
halo star with sufficient S/N ($\gtrsim{15}\,$\AA$^{-1}$) 
for chemical abundance analysis. Four of these candidates have a 
high likelihood of being \andromeda\ halo stars (Section~2.1).  The other 
six stars are marginally identified as \andromeda\ halo stars.  The 
likelihood distributions of MW foreground stars and \andromeda\ halo 
stars overlap, and the number of MW foreground stars in the sample is 
large at the distances from \andromeda's center 
considered here (Figure 3 of G12).  Therefore, we expect contamination 
from MW foreground stars among the stars that are marginally identified 
as \andromeda\ stars.  For these reasons, we restrict our sample to the four 
high likelihood \andromeda\ members.  

\subsection{Characteristics of the Halo Sample}\label{ssec_sample}

The \nstars\ stars in our sample are located at projected distances 
from \andromeda's center of $R_{\text{proj}}$\,$\sim70$ to 140 kpc.  
At these distances, the contribution from an in-situ halo component is 
expected to be negligible. Instead, the halo should be dominated by 
metal-poor stars belonging to an accreted component \citep{Zolotov2012a}.

Due to the proximity of the sample stars to various \andromeda\ 
satellites, we check whether our stars could be members of the 
nearby satellites. In the top panels of Figure~\ref{fig: membership}, 
we plot the relative location in position-velocity space between each 
program halo star and the nearest satellite. The plots also include 
other high-likelihood halo stars (G12) as well as satellite member 
stars \citep{Ho2012a,Tollerud2012a}. The halo stars fall far from 
the locus of satellite member stars. In the bottom panels, we 
show CMDs for the same data as in the top row. Due to the 
similar line-of-sight distances to \andromeda\ and the satellites, 
CMDs are not a good discriminant of halo versus satellite 
membership. 

\begin{figure}[tpb!]
\centering
\includegraphics[width=.49\textwidth]{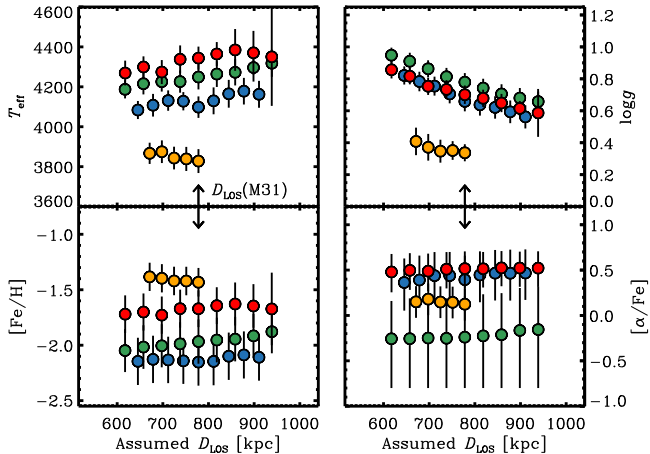}	
\caption{Variation of \teff,  \logg, \feh, and \afe\, for a range of 
assumed line-of-sight distances, using the colors in Figure~\ref{fig: membership}. 
The arrows mark the distance to \andromeda. The yellow star is not 
plotted for $D_{\text{LOS}} > 780$ kpc as at those distances it would 
be brighter than the tip of the RGB. See Section~\ref{ssec_distances} 
for further discussion.}
\label{fig: distance_parameters}
\vskip 2pt
\end{figure}

\begin{figure*}[tpb!]
\centering
\includegraphics[width=.76\textwidth]{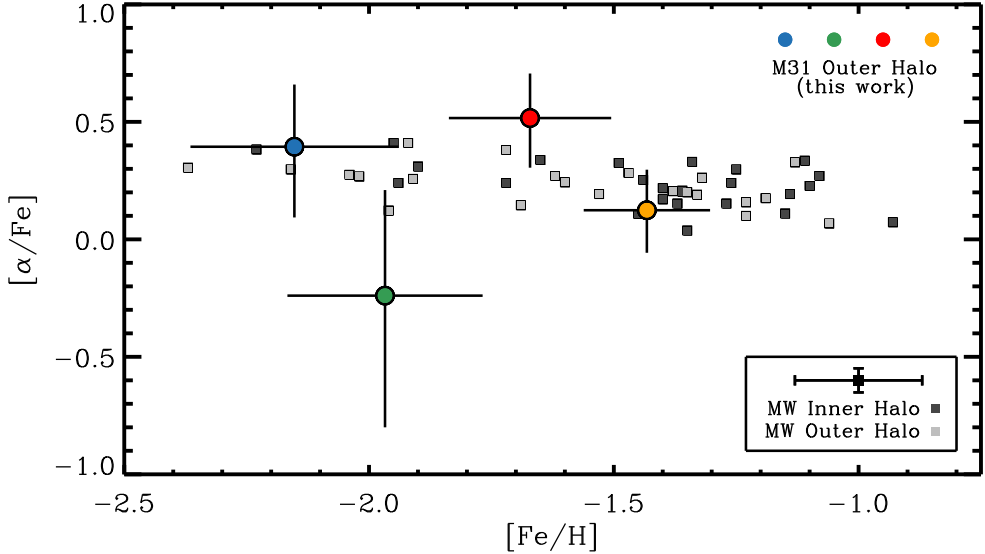}	
\caption{\afe\ as a function of \feh\, for the four \andromeda\, halo 
stars. Small squares indicate MW inner/outer halo stars 
measured with high resolution by \citet{Ishigaki2012a}. 
A representative error bar for the MW halo measurements 
is included in the legend. The \andromeda\ halo stars are 
alpha-enhanced on average, similar 
to that in Milky Way halo stars of comparable metallicities.}
\label{fig: alphafe_feh}
\end{figure*}

\section{Chemical Abundance Analysis}\label{sec_analysis}

For the \nstars\ \andromeda\ halo stars above, we measure iron 
abundances, \feh, and alpha to iron abundance ratios, 
\afe, by comparing each DEIMOS spectrum to a large 
grid of synthetic spectral models \citep{Kirby2011c}. 
The technique has been applied to the study of MW
satellites \citep{Kirby2011b,Vargas2013a} and M\,31 
satellites (V14). We focus on those aspects most pertinent 
to this work, and refer the reader to V14 for a detailed 
method description.

\subsection{Measuring \feh\ and \afe}\label{ssec_analysis}

To find the best-fitting synthetic model, 
we use wavelength regions sensitive to Fe or alpha 
elements (here Mg, Si, Ca and Ti) and minimize the 
$\chi^2$ flux difference between the spectrum and the
grid models. The synthetic grid samples a wide range 
of \teff, \logg, \feh\, and \afe. After measuring the best-fitting
abundances, we apply a correction factor to \afe\ derived 
by V14 so that \afe\ better represents an unweighted 
average of [Mg/Fe], [Si/Fe], [Ca/Fe], and [Ti/Fe]. 

Due to the large distance to \andromeda, our spectra have relatively 
low S/N. Using synthetic spectra,  V14 
confirmed that our method can measure both alpha-solar 
($\sim+0.0$) and alpha-rich ($\sim+0.4$) values in  
luminous RGB stars using S/N $\gtrsim$ 15 \AA$^{-1}$ 
spectra with more than 93\% completeness (see Figure~3 of  V14). 

\subsection{Lack of Distances to \andromeda\, Halo Stars}\label{ssec_distances} 

In V14, we fixed \logg\ from the best color and magnitude 
fit to isochrones shifted to the line-of-sight distance, $D_{\text{LOS}}$,
of each galaxy. For field halo stars, $D_{\text{LOS}}$ is loosely 
constrained by the assumption that the stars lie within the \andromeda\ stellar halo. 
To test the change of abundance with assumed distance, we perform the 
abundance analysis independently for nine assumed distances, ranging 
between $D_{\text{\andromeda}} \pm \Delta$$D$ (see lower panels of 
Figure~\ref{fig: membership}). $D_{\text{\andromeda}}$ is 
the distance to \andromeda\ (779 kpc; \citealt{Conn2012a}), and $\Delta$$D$ 
is the line-of-sight distance between \andromeda\ and the outskirts of the 
stellar halo at the projected distance to each star from \andromeda. We 
assume a spherical stellar halo\footnote{The shape of the halo may 
deviate from sphericity in its inner regions, but all of our stars are at least 
70 kpc away from \andromeda.} with radius $R_{\text{halo}}=175$ kpc, 
equal to the projected distance to \andromeda\  of the most distant halo 
stars securely identified by G12.  

We show the results of this test in Figure~\ref{fig: distance_parameters}. 
The top panels show the variation in \teff\ and \logg\ with
assumed $D_{\text{LOS}}$. For larger $D_{\text{LOS}}$, 
\teff\ tends to increase (slightly) while \logg\ decreases. 
The lack of significant variation in \teff\ is due to the primary 
dependence of \teff\ on color and not on luminosity. 
The variation in \logg\ is due to the star occupying a 
higher-luminosity position in the CMD for larger assumed 
$D_{\text{LOS}}$. For the yellow star, stellar parameters 
are only shown for $D_{\text{LOS}}\leq$ $D_{\text{\andromeda}}$;
for larger $D_{\text{LOS}}$, the star lies above all 
isochrones, and would instead have to be considered 
a TP-AGB star. The number of TP-AGB stars is highly 
model dependent, but observations with large AGB 
samples suggest that for low metallicity systems, the 
fraction of TP-AGB relative to RGB stars is less than 
a few percent \citep{Girardi2010a}. Thus, we suggest that 
it is improbable that this star is a TP-AGB member.

The bottom two panels show how changes in $D_{\text{LOS}}$ 
translate to changes in \feh\ and \afe. All abundance 
measurements vary by less than  $\sim0.15$ dex, a
change equal or smaller than all $1\sigma$ measurement 
uncertainties. We note  that due to the centrally 
concentrated \andromeda\ halo it is most likely that each 
star will have a line-of-sight distance similar to that of \andromeda.

Another source of uncertainty is stellar age, since stellar
parameters are measured using a grid of 12\,Gyr isochrones. 
We test for the dependence on isochrone age by using younger,
4\,Gyr isochrones. The changes in \feh\ and \afe\ are 
$-0.02$ and $-0.06$ dex, respectively, lower than 
our minimum measurement uncertainties. Given the lack of 
significant change in abundances, we assume $D_{\text{LOS}}$ 
= $D_{\text{\andromeda}}$, and use only 12\,Gyr old isochrones
for the analysis below.

\section{Results}\label{sec_results}

We have measured \feh\, and \afe\, for \nstars\ outer halo stars of 
\andromeda\ spanning $\sim$1 dex in \feh. We measure \feh\ $=$  $-2.15\pm{0.21}$, $-1.97\pm{0.20}$,
$-1.67\pm{0.17}$, $-1.43\pm{0.13}$; and \afe = $+0.39^{+0.26}_{-0.30}$, $-0.23^{+0.45}_{-0.56}$, 
$+0.52^{+0.19}_{-0.21}$, $+0.12^{+0.17}_{-0.18}$).The stars 
are located between $R_{\text{proj}}\sim{70}$ and 140 kpc from \andromeda. 
In Figure~\ref{fig: alphafe_feh} we plot our \feh\, and \afe\, measurements.   
Three out of the four stars have \afe\, $> 0.0$, consistent with typical MW 
halo values within the uncertainties (see
Section~\ref{ssec_average}). While the fourth star has \afe\ $<$ 0, its large uncertainty does not allow 
us to distinguish whether it is drawn from an alpha-poor population 
enriched by Fe-rich Type~Ia SNe, or is an \afe\ enhanced star scattered 
to low \afe. Low \afe\ and \feh\ stars have been detected in the MW halo 
but are rare \citep{Ivans2003a}. Our sample is too small to detect an intrinsic range 
in \afe\ in the \andromeda\ halo.

\subsection{Average \afe\ in the M\,31 Outer Halo and Comparison to Milky Way}\label{ssec_average}

\begin{figure}[tpb!]
\centering
\includegraphics[width=.49\textwidth]{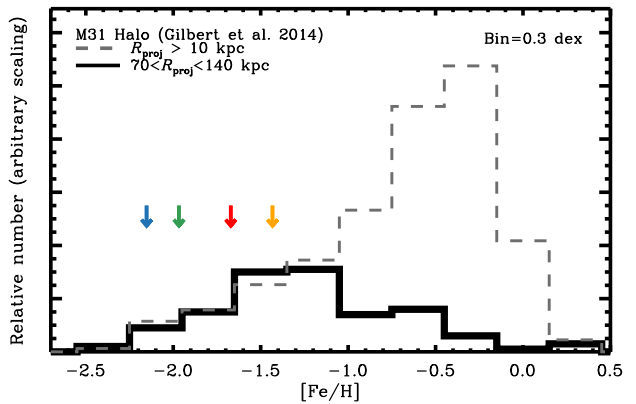}	
\caption{MDF for \andromeda\ outer halo stars in the same range 
of distances as our \afe\ sample (solid histogram), and for halo stars 
at all distances (thin dashed histogram), using the photometric metallicities 
from G14. The bin size is equal to the mean \feh\ uncertainty for metal-poor stars 
in the MDF ($=\sigma\sim{0.29}$). The y-axis is scaled arbitrarily to show the shapes of the two histograms. The four stars 
presented in this paper span roughly the metal-poor half of the outer 
halo MDF.}
\label{fig: m31halo_mdf}
\end{figure}

To draw conclusions from our small sample, we determine the 
sample average \afe\ ratio in two ways. First, we measure the unweighted 
average and its uncertainty taking into account the small sample size from 
the Student-\textit{t} distribution. We obtain \afeavg$=+0.20\pm0.20$.
An inverse-variance weighted mean yields \afeavg$=+0.28\pm0.12$, 
but Monte Carlo tests by V14 suggest that this weighting could 
bias the average high by $\sim+0.05-0.10$ due to a modest 
anti-correlation between \afe\ and its uncertainty. Second, 
we recalculate \afeavg\ by weighting our measurements by 
their expected fractional contribution to the halo from \citet{Gilbert2014a}'s 
metallicity distribution function (MDF), shown in Figure~\ref{fig: m31halo_mdf}. 
We use the thicker histogram, corresponding to stars at the 
same projected distances from \andromeda\ as our stars. We 
draw 10,000 realizations of \feh\ and \afe\ from each of our four 
stars using their best-fit values and individual errors. Each 
realization is thus a sample of four (\feh, \afe) pairs. We reorder 
the pairs by \feh\, and calculate weights for each \afe\ value 
as the area under the normalized MDF bounded by the midpoint 
between the \feh\ value of each of our data points and 
the previous/next value. We truncate the MDF to the range 
of \feh\ abundances of our stars. From the 10,000 realizations, 
we measure \afeavg$=+0.19\pm{0.14}$, in agreement 
with the simple average.

We next compare \afe\ between the MW and \andromeda\ halos. 
For the MW, we use the homogeneous sample by 
\citet{Ishigaki2012a}, who classify stars as inner or 
outer  halo stars kinematically. We select all 
68 stars with reported measurements of [Mg/Fe], [Si/Fe], 
[Ca/Fe], and [Ti/Fe] to calculate \afe\ consistently with our 
own definition of \afe\ (see Section~\ref{ssec_analysis}). The 
MW points are also plotted in Figure~\ref{fig: alphafe_feh}. 
We calculate \afeavg\ for both MW inner and outer halo 
subsamples, finding unweighted averages of \afeavg$ = +0.23$ 
in both cases. Thus, our average \afe\ value for \andromeda's outer 
halo stars (\afeavg $= 0.20\pm{0.20}$) is consistent with the MW 
value within the uncertainties (in the same range of metallicities).

\subsection{M\,31 Halo versus Satellite Galaxies}\label{ssec_dwarfs}

\begin{figure}[tpb!]
\centering
\includegraphics[width=.48\textwidth]{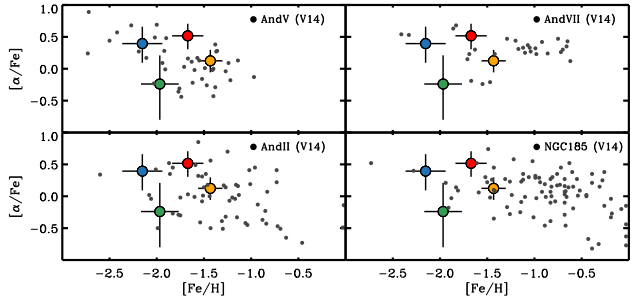}	
\caption{Comparison of the four field halo star \afe\ abundances with 
those of the four \andromeda\ satellites with the largest number of 
measurements (V14, small circles).  Both halo and satellite samples 
come from DEIMOS spectra of RGBs with similar S/N, analyzed with 
the same method. The halo sample is more metal-poor than the two 
brighter satellites.}
\label{fig: alphafe_feh_dwarfs}
\end{figure}

The lower \afe\ ratios in MW surviving satellites relative to
MW halo stars \citep[e.g.,][]{Venn2004a,Kirby2011b} have been 
used to infer that MW halo progenitors had short-lived and rapid star 
formation histories relative to the surviving satellites \citep{Robertson2005a}. 
Using our sample and V14, we compare in Figure~\ref{fig: alphafe_feh_dwarfs} 
the alpha abundances between the halo and the four present-day 
\andromeda\ satellites with the largest samples (And~V, And~VII, 
And~II, and NGC~185). The satellites range in luminosity from 
$7\times10^{5}$ $L_{\odot}$ (And~V) to $1.8\times10^{8}$ 
$L_{\odot}$ (NGC~185). Thus, they sample a wide range
of the satellite luminosity function except for the faint end.

The average \feh\ of our halo sample is
most similar to those of the less luminous And~V and And~VII, 
whereas And~II and NGC~185 are more metal-rich. Our small 
sample size precludes strong statistical comparisons between 
halo and satellite abundances, but we make a first attempt by 
comparing \afeavg. V14 measured average \afe\ ratios of 
$0.12\pm0.09$, $0.30\pm0.09$, $0.03\pm0.09$ and 
$0.12\pm0.09$ for And~V, And~VII, And~II, and NGC185, 
respectively.  Thus, none of the satellites have an \afeavg\ value 
discrepant by more than 1$\sigma$ from the halo value 
calculated in Section~\ref{ssec_average}, \afeavg$=+0.20\pm0.20$,
The comparison between the halo and satellites must thus be 
revisited with larger samples.

\section{Discussion and Conclusions}\label{sec_disc}

We present the first \afe\ measurements of stars in 
a stellar halo beyond the MW. The average \afe\ of our \nstars\ 
\andromeda\ outer halo stars is $0.20\pm{0.20}$. The 
metallicities of the \nstars\ stars sample roughly the metal-poor
half of the metallicity distribution of the outer halo at distances 
comparable to those of our \nstars\ stars ($70 \lesssim$\,$R_{\text{proj}}\lesssim 140$ kpc). 

High alpha enhancements are characteristic of Type~II SNe 
enrichment of the ISM operating on short timescales. Thus, 
the best-fit high average \afe\ ratio in the outer \andromeda\ 
halo combined with the low halo metallicity at these projected 
distances hints towards a progenitor population that came into 
proximity with the proto-\andromeda\ halo at early times. In 
simulations of halo formation, dwarf satellites with high \afeavg\ 
and low \fehavg\ correlate with elevated star formation rates, 
early accretion, \citep{Johnston2008a,Tissera2012a} 
and/or were more affected by gravitational tidal interactions \citep{Nichols2014a}. 
An alternate pathway for producing a high \afe\ population is to 
invoke the dissolution of globular clusters. In the MW halo, globular 
clusters may have contributed anywhere from a few to $50$\% of the 
halo stellar mass \citep[e.g.,][]{Carretta2010c,Martell2011a}. Given 
the \afe\ enhancements of present-day \andromeda\ globular 
clusters (\afeavg$=+0.37\pm{0.16}$, \citealt{Colucci2009a}), this formation 
pathway may also hold in the \andromeda\ halo.

The \andromeda\ halo enrichment pattern agrees with that of 
the MW halo, whether considering inner or outer halo stars. 
Given the apparent differences in accretion histories between 
the MW and \andromeda\ stellar halos, it is somewhat surprising 
that both share similar chemical properties (at similar metallicities). 
More measurements of \afe\ in outer halo stars of the MW and \andromeda\ 
are needed to better compare the chemical properties of the 
principal progenitors of both outer halos.

\section*{Acknowledgements}

LCV acknowledges useful  
conversations with Nikhil Padmanabhan during
the analysis stage of this project. LCV was supported by the National Science 
Foundation Graduate Research Fellowship (Grant No. \mbox{DGE$-$1122492}).  MG and 
LCV acknowledge support from NSF Grant \mbox{AST$-$0908752}, 
and PG acknowledges support from NSF grant \mbox{AST$-$1010039}. 
EJT and KG were supported by NASA through Hubble Fellowship Grant 
Nos. 51316.01 and 51273.01, awarded by the Space Telescope Science 
Institute, which is operated by the Association of Universities for Research 
in Astronomy, Inc., for NASA, under contract NAS 5-26555. \textit{Facilities:} 
Keck II (DEIMOS)

\bibliographystyle{apj}

\end{document}